\documentclass[twocolumn,axodraw]{article}
\usepackage{epsfig,axodraw}
\input epsf.tex

\def\beq{\begin{eqnarray}}
\def\eeq{\end{eqnarray}}

\let\badcite=\cite
\def\cite{~\badcite}

\def\slashchar#1{\setbox0=\hbox{$#1$}           % set a box for #1
   \dimen0=\wd0                                 % and get its size
   \setbox1=\hbox{/} \dimen1=\wd1               % get size of /
   \ifdim\dimen0>\dimen1                        % #1 is bigger
      \rlap{\hbox to \dimen0{\hfil/\hfil}}      % so center / in box
      #1                                        % and print #1
   \else                                        % / is bigger
      \rlap{\hbox to \dimen1{\hfil$#1$\hfil}}   % so center #1
      /                                         % and print /
   \fi} 
    \def\slashword#1{\setbox0=\hbox{$#1$}        %     set a box for #1
  \dimen0=\wd0                                   %and get its size
   \setbox1=\hbox{/} \dimen1=\wd1                % get size of /
   \ifdim\dimen0>\dimen1                         % #1 is bigger
      \rlap{\hbox to \dimen0{\hfil\bf---\hfil}} %        
      #1                                         %
   \else                                         % / is bigger
      \rlap{\hbox to \dimen1{\hfil$#1$\hfil}}    % so center #1
      /                                          % and print /
    \fi}                                         %

\catcode`@=11
\newdimen\vbigd@men                             % for \vbig

\def\vbig#1#2{{\vbigd@men=#2\divide\vbigd@men by 2%
   \hbox{$\left#1\vbox to \vbigd@men{}\right.\n@space$}}}
\catcode`@=12

\catcode`@=11
\def\citenum#1{\csname b@#1\endcsname}
\catcode`@=12

\title{The GZK Bound in Discrete Space}

\author{{\bf N. Kersting}\\Department of Physics\\Sichuan University\\ Chengdu, 610065 P.R.
 China\\{\normalsize \tt kest@mail.tsinghua.edu.cn}}
\date{}

\begin{document}

\maketitle

\begin{flushleft}
{SCUPHY-TH-06001}\\
\end{flushleft}

\begin{abstract}

{\it The maximum distance bound for ultrahigh energy cosmic rays (UHECR) with energies above the 
Greisen-Zatsepin-Kuzmich cutoff $\sim 10^{19}~eV$ 
relaxes significantly if there is some mechanism that forces UHECR to propagate 
in discrete intervals, rather than continuously, through intergalactic space. In particular,
intervals as small as a femtometer relax the bound for protons by an order of magnitude and
potentially account for the observed excess of UHECR flux.
}
\end{abstract}

%\newpage
%\pagestyle{empty}
%\end{titlepage}

%%%%%%%%%%%%%%%%%%%%%%%%%%%%%%%%%%%%%%%%%%%%%%%%%%%%%%%%%%%%%%%%%%%%%%
\section*{Introduction}
\label{sec:intro}

Recent experiments\cite{ex1, ex2, ex3} have indicated that the flux of cosmic rays
entering the Earth's atmosphere
with energy beyond the Greisen-Zatsepin-Kuzmich (GZK) limit\cite{GZK}
of roughly $5 \cdot 10^{19}~eV$ is much higher than expected.
According to the original papers on the subject, such ultra-high-energy 
cosmic rays (UHECR), presumably consisting of photons, nucleons, or heavier
nuclei, must be rapidly attenuated after travelling a distance of 
$O(10)$ Mpc through intergalactic space due to interactions such as
$N+\gamma \to \Delta \to N + \pi$ and $\gamma + \gamma \to e^{+} e^{-}$
 with the cosmic
microwave background radiation (CMBR), and since there are no known sources
of UHECR within this radius of the Earth the observed flux in our atmosphere should be negligable. 
If further experimentation verifies excess in UHECR flux then the simple
GZK model may not be the correct theory of UHECR propagation in space.

The excess of UHECR has been known for at least a decade\cite{uhecr-old}, but of interest lately is analysis 
  showing significant clustering of UHECR 
events in the sky in directions that point\footnote{These directions are meaningful because  intergalactic magnetic fields 
typically bend the
paths of UHECR by less than a degree\cite{bend}} to candidate
sources $~140$ MpC away\cite{source140}. The possibility that UHECR are able to travel such long distances through the
CMBR lends strong motivation to revise the GZK bound.

The original derivation of the GZK bound for protons depends on the energy of the protons (E), the
energy loss per interaction ($\Delta E$) with a CMBR photon, the number density of CMBR 
photons (n), and the interaction cross-section ($\sigma$) as follows:
\beq
\label{lgzk}
\lambda_{GZK} = (\frac{E}{\Delta E})(\frac{1}{n \sigma})
\eeq
For example, for $E=10^{20}~eV$ protons incident on CMBR photons giving
 resonant pion-production we have $\sigma \approx 200~\mu b$, 
$n \approx 400 ~cm^{-3}$,  and therefore $\lambda_{GZK} \approx 10^{24}~m \approx 30~Mpc$ (one can
derive a similar limit for photons).
The dilemma is that no potential UHECR sources, such as Active Galactic Nuclei,
are observed to be within that radius of the Earth\cite{sources}.

Models to resolve the dilemma usually introduce additional assumptions to 
modify  $\Delta E$ or  $\sigma$ in (\ref{lgzk}), resulting in a larger  $\lambda_{GZK}$.
For example,  $\Delta E$ can decrease in theories in which energy conservation fails at high energies \cite{lorentz}; $\sigma$ decreases
 in models where the UHECR is
 initiated by an exotic
particle or neutrino which interacts very weakly with the CMBR\cite{exotic, neutrino}.

The current Letter proposes that one can extend the GZK bound if
UHECR propagate in
discrete jumps through space-time, rather than in a continuous path.
It turns out that the form of the second factor in (\ref{lgzk}) changes, 
and a jump size as small
as a few femtometers is sufficient to resolve the dilemma.
In contrast to the above-metioned theories, this model introduces no new
particles or violation of conservation principles, but rests merely on
the hypothesis that space-time is discrete.

\section*{Calculation}

First let us recall the derivation of the second factor in parenthesis 
in (\ref{lgzk}), the mean free path length for, say, a proton. This comes about from a 
pseudo-classical
approximation of the proton-CMBR interaction, justified by the relative diffuseness of
the CMBR.
As shown in Figure 1, from the perspective
of the CMBR the proton
sweeps out an effective tube in 3-space of cross-sectional area $\sigma$ and length $R$;
 the probability of 
interaction with a CMBR photon along this path is just the volume swept out,
 $\sigma R$, divided by the photon
number density $n$. Therefore the probability of interaction per unit path length
is constant, and its inverse, the mean free path of the proton, is $(n\sigma)^{-1}$.

{
\unitlength=1.3 pt
\SetScale{1.25}
\SetWidth{0.5}      % line    size control
\scriptsize    %  letter  size control
% 1
\begin{picture}(90,50)(0,10)
\Line(30,20)(110,20)
\Line(30,40)(110,40)
\Oval(30,30)(10,5)(0)
\Oval(110,30)(10,5)(0)
\Text(29,30)[c]{\large $\sigma$}
\Text(70,10.5)[c]{\large $R$}
%\SetPFont{roman}{5.5}   
%\PText(70,10)(90)[c]{\large $\}$}
\end{picture} \
}
\begin{center}
Figure 1.
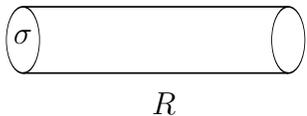
{\it Segment of UHECR path}
\end{center}

Now suppose the proton doesn't move in a continuous line, but propagates in 
jumps\footnote{Observable properties of the proton such as velocity, energy, etc. are then
defined as
suitable averages over many jumps.} 
of size $R$. Working in the same pseudo-classical approximation above,
 in the course of one jump the proton occupies an approximate
 volume of
$\sigma^{3/2}$ in the CMBR, before
jumping again. The mean path length (though this 'path' is traced out over many jumps)
 is then $ \frac{R}{ n \sigma^{3/2}}$. Since 
the fractional energy lost per interaction remains the same, the GZK bound should be replaced
by
\beq
\label{lrgzk}
\lambda_R = (\frac{E}{\Delta E})(\frac{ R}{ n \sigma^{3/2}})
\eeq
and depends on the size $R$ of the proton jumps. 
Note that as $R \to \sqrt{\sigma}$, $\lambda_R$ becomes $\lambda_{GZK}$, and
for  $R < \sqrt{\sigma}$  $\lambda_{GZK}$ replaces  $\lambda_R$.

Using again the example of photopion production, with  $R \approx 1~fm$,
 gives $\lambda_R = 300~MpC$.
Remarkably, if UHECR protons jump in intervals as small as a femtometer
they can easily traverse the intergalactic space to the nearest UHECR sources $140$ MpC
away.

\section*{Comments}

Supposing UHECR protons do jump in fermi-size steps through intergalactic
space, an immediate concern is whether  this behaviour applies to 
 other energies and other particles: i.e. few researchers would object to 
discreteness at the Planck-scale, but do
existing observations rule out spacetime discreteness at the fermi-scale?
One expects physics of very low energies (eV- and below) 
to be insensitive to motions on the order of a fermi.
However, already in the realm of medium- to high-energies($E>MeV$) severe constraints may
arise in more sensitive environments, e.g. stellar interiors or particle detectors in
accelerator-based experiments,
though these
are not nearly as uniform and diffuse as the CMBR and hence require a more sophistcated
treatment than that appearing in this Letter.

In the event, however, in which the discretization of space is a dynamic
phenomenon depending on energy (and possibly other quantum numbers) these
phenomenological constraints may be weakened or removed, and investigations
in this direction are underway\cite{wip}.

\end{document}